\documentclass[final,5p,times,twocolumn]{elsarticle} 
 
\usepackage{amssymb} 
\usepackage{amsmath} 
\usepackage{hyperref}
\hypersetup{
  colorlinks=true,
  linkcolor=blue,
  citecolor=blue,
  urlcolor=blue
} 
\usepackage{mathrsfs}

\usepackage{booktabs}
\usepackage{graphicx}

\journal{Physica A: Statistical Mechanics and its Applications}

\begin{document}

\begin{frontmatter} 

\title{Forward-Oriented Causal Observables for Non-Stationary Financial Markets}

\author[aff1]{Lucas A.~Souza\corref{cor1}}
\cortext[cor1]{Corresponding author}
\ead{lasouza@if.usp.br}

\affiliation[aff1]{
  organization={S\~ao Paulo},
  % city={S\~ao Paulo},
  state={SP},
  country={Brazil}
}

\begin{abstract}
We study short-horizon forecasting in financial time series under strict causal constraints, treating the market as a non-stationary stochastic system in which any predictive observable must be computable online from information available up to the decision time.
Rather than proposing a machine-learning predictor or a direct price-forecast model, we focus on \emph{constructing} an interpretable causal signal from heterogeneous micro-features that encode complementary aspects of the dynamics (momentum, volume pressure, trend acceleration, and volatility-normalized price location).
The construction combines (i) causal centering, (ii) linear aggregation into a composite observable, (iii) causal stabilization via a one-dimensional Kalman filter, and (iv) an adaptive ``forward-like'' operator that mixes the composite signal with a smoothed causal derivative term.
The resulting observable is mapped into a transparent decision functional and evaluated through realized cumulative returns and turnover.
An application to high-frequency EURUSDT (1-minute) illustrates that causally constructed observables can exhibit substantial economic relevance in specific regimes, while degrading under subsequent regime shifts, highlighting both the potential and the limitations of causal signal design in non-stationary markets.
\end{abstract}




\end{frontmatter}

%\tableofcontents



% =========================================================
\section{Introduction and Related Literature}
\label{sec:intro} 
Financial markets can be viewed as complex stochastic systems, where collective dynamics, information flow, and feedback effects give rise to non-trivial temporal correlations and non-stationary behavior, a perspective long explored in statistical physics and nonlinear dynamics \cite{schreiber2000,marschinski2002}.
Within this framework, price dynamics emerge from the interaction of heterogeneous agents and exhibit noisy, nonlinear, and time-varying features that challenge equilibrium-based descriptions.
The application of tools from statistical physics to financial markets has revealed well-documented empirical regularities, including heavy-tailed return distributions, volatility clustering, and excess kurtosis \cite{bouchaud2003theory, cont2000herd, mantegna1995}.
Such emergent collective effects deviate markedly from the Gaussian benchmark and motivate modeling approaches that explicitly account for feedback mechanisms, clustering, and non-stationarity.

A central modeling question concerns the origin of large price moves and volatility bursts: to what extent are they driven by exogenous information shocks (e.g., macroeconomic news) versus endogenous dynamics arising from internal amplification mechanisms?
Recent empirical evidence supports a clear dynamical distinction between exogenous and endogenous jumps.
Exogenous events typically manifest as abrupt volatility spikes followed by power-law relaxation reminiscent of Omori's law for earthquake aftershocks, whereas endogenous events are often preceded by a gradual build-up of volatility, consistent with self-exciting feedback mechanisms
\cite{Lallouache2020,Marcaccioli_2022}.
This perspective naturally connects to the broader literature on endogenous market activity and reflexivity, in which self-exciting point processes—most notably Hawkes processes—provide a parsimonious mathematical framework to represent event clustering and cascading behavior \cite{hawkes1971}.

In parallel, the analysis of causality and information flow across market variables has become an important tool for studying short-horizon predictability in noisy environments.
Nonlinear dependence measures such as transfer entropy offer a principled way to quantify directed information flow between observables beyond linear correlation \cite{schreiber2000,marschinski2002}.
Building on these ideas, recent forecasting pipelines have combined causality-based covariates with modern sequence models to improve predictive performance in multivariate financial settings \cite{berenguer2024}.
Moreover, even at the microstructural level, physical constraints on information transmission impose irreducible limits on simultaneity across geographically separated trading venues, with direct implications for high-frequency trading and market regulation \cite{Angel2014physicsAndFinancial}.

Against this background, a strict practical constraint is often under-emphasized: any tradable predictive observable must be computed \emph{online}, using exclusively information available up to the decision time.
In practice, seemingly innocuous steps—such as smoothing, normalization, or label and target construction—can introduce subtle look-ahead biases that materially inflate backtest performance.
While recent work has increasingly employed machine learning and statistical learning techniques to extract patterns from financial time series
\cite{zahedi2015,rounaghi2015,roostaee2023,wood2024},
robustness under regime shifts remains a persistent challenge, including well-documented breakdowns of trend and momentum effects during market stress episodes \cite{daniel2016}.

Motivated by the need for signals that preserve strict causal validity while remaining economically meaningful, this paper focuses on engineering an interpretable composite observable from heterogeneous technical features, including momentum, volume pressure, trend acceleration, and volatility-normalized price location.
Rather than performing direct price prediction, we construct a causal signal stabilized via a one-dimensional Kalman filter~\cite{kalman1960} and introduce an adaptive forward-like operator that combines the filtered signal with a smoothed causal derivative component.
The resulting signal is evaluated through a transparent threshold-based decision rule applied to high-frequency EURUSDT (1-minute) data.

Empirically, we find that the proposed causal signal exhibits substantial economic relevance over specific market regimes.
However, its effectiveness is strongly regime-dependent and degrades following structural changes in the underlying dynamics.
These findings reinforce the non-stationary and adaptive nature of financial markets and motivate the methodological focus of this study on causal signal construction, filtering, and regime sensitivity under strict information constraints.

% =========================================================
\section{Methodology} 

This section formalizes the causal forecasting setup adopted in this study.
All signals, transformations, and decision rules are constructed under strict causal constraints, relying exclusively on information available up to time $t$.
The focus is on signal design and causal filtering rather than on explicit price prediction.

\subsection{Forecasting Under Causal Constraints}
\label{subsec:causal_constraints}

Under strict causality, forecasting at short horizons is constrained by the limited information content available at the decision time.
Signals must be constructed from contemporaneous and past observations only, while remaining sufficiently stable to support decisions and sufficiently responsive to reflect regime changes.
This tension between responsiveness and stability constitutes a central challenge in causal signal design.

A second challenge arises from the use of technical indicators and filtering operations.
Many standard preprocessing steps---such as smoothing, normalization, or target construction---implicitly rely on future information when applied naively.
In a causal setting, these operations must be reformulated so that no information beyond time $t$ enters the signal at any stage of construction.

Finally, economic relevance imposes an additional constraint.
A causally valid signal may exhibit appealing statistical properties while failing to translate into meaningful realized performance under a decision rule.
For this reason, the methodology emphasizes constructions that can be mapped into transparent decision functionals and evaluated through realized outcomes (returns and turnover), rather than relying solely on forecast error metrics.

\subsection{Causal Signal Construction}
\label{subsec:causal_signal}

We construct a predictive observable by aggregating heterogeneous sources of market information in a fully causal manner.
Rather than relying on a single indicator, we combine representative measures of momentum, volume pressure, trend acceleration, and volatility-normalized price location.

Each component captures a distinct informational dimension: relative strength (RSI), volume-adjusted momentum (MFI), trend acceleration (MACD difference) and volatility-normalized price position (BB \%).
The indicators are not used as standalone trading rules, but as features entering a unified signal construction. Throughout the analysis, all indicator lookback windows and parameter ranges are fixed to conventional values widely adopted in the technical analysis literature and in market practice, with no optimization performed.

To ensure comparability across indicators and avoid look-ahead bias, each series is centered using a causal median operator:
\begin{equation}
\tilde I_t^{(k)} = I_t^{(k)} - \operatorname{median}\{ I_\tau^{(k)} : \tau < t \}.
\end{equation}

The centered indicators are linearly combined into a raw composite signal:
\begin{equation}
\mathscr{F}_0(t) = \frac{1}{K} \sum_{k=1}^K \alpha_k \tilde I_t^{(k)}.
\end{equation}
As the present study employs four indicators, the aggregation takes the explicit form
\begin{equation}
\mathscr{F}_0(t)
 = \frac{1}{4}\Big(
\alpha_{\text{MFI}} \, \tilde I_t^{(\text{MFI})}
+ \alpha_{\text{RSI}} \, \tilde I_t^{(\text{RSI})}
+ \alpha_{\text{BB}\%} \, \tilde I_t^{(\text{BB}\%)}
+ \alpha_{\text{MACD}} \, \tilde I_t^{(\text{MACD})}
\Big),
\end{equation}
where the coefficients $\alpha_k$ are fixed scaling constants chosen to account for the typical magnitude of each indicator.

\begin{figure*}[ht]
\centering
\includegraphics[width=0.8\textwidth]{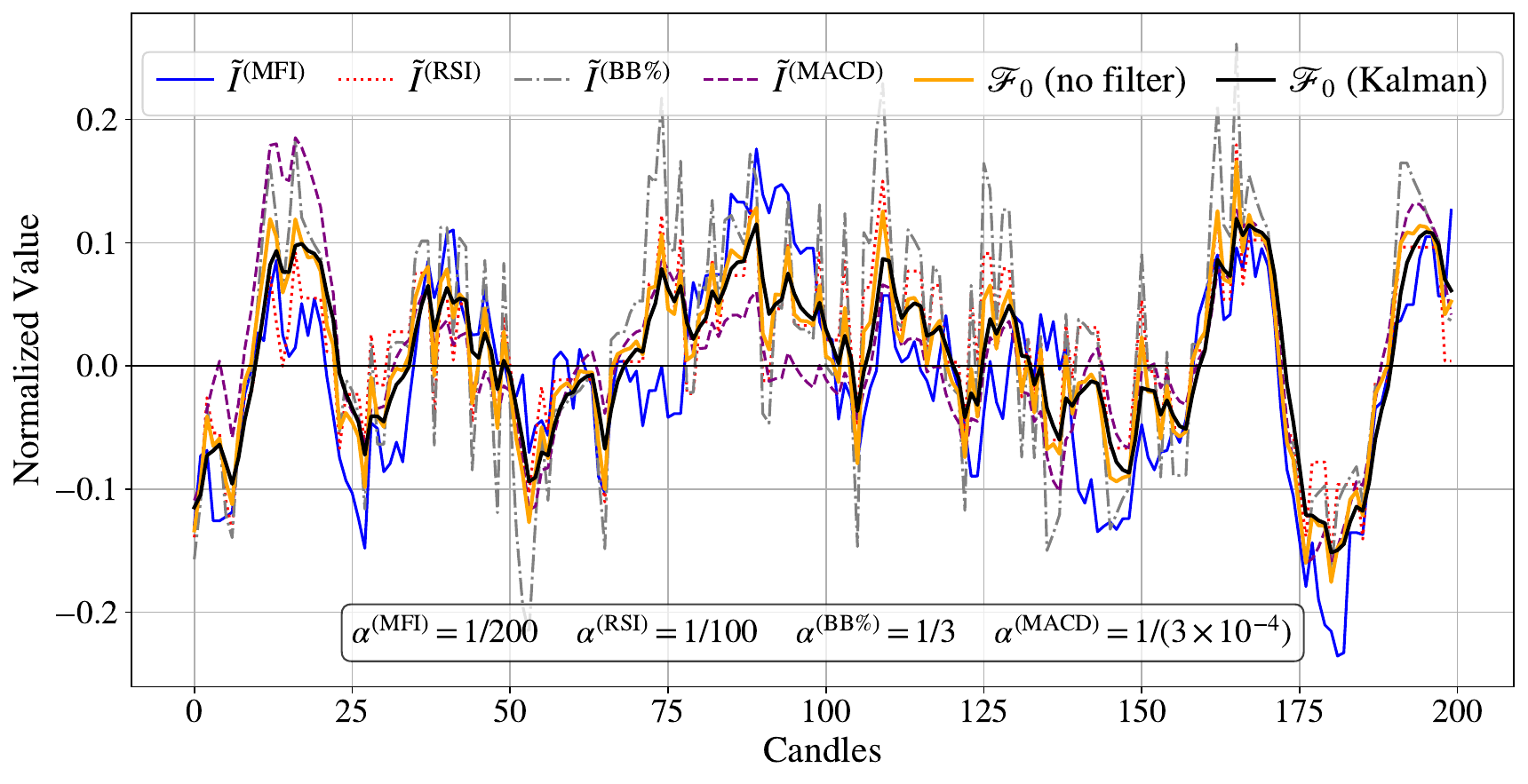}
\caption{Causal construction of the composite observable. The causally centered indicators are linearly aggregated into the raw composite signal $\mathscr{F}_0$ and subsequently smoothed via a one-dimensional Kalman filter. The linear scaling coefficients ($\alpha$) are indicated in the figure.}
\label{fig:indicators}
\end{figure*}

Figure~\ref{fig:indicators} illustrates the intermediate stages of the causal signal construction using a representative 1-minute EURUSDT excerpt.
The figure displays the indicator components after causal centering and the resulting composite $\mathscr{F}_0(t)$.
Because the raw aggregation remains sensitive to high-frequency fluctuations, a causal smoothing step based on a one-dimensional Kalman filter is applied, yielding a stabilized indicator-level observable that serves as an intermediate building block for the final predictive signal.
The Kalman filter parameters are fixed to $q = 0.01$ and $r = 0.1$; further implementation details are provided in~\ref{app:kalman}.
The dependence of the resulting signal on the Kalman filter parameters is acknowledged, but not explored systematically in this work.

\subsection{Approximate Forward Operator}
\label{subsec:forward_operator}

\begin{figure*}[ht]
\centering
\includegraphics[width=0.8\textwidth]{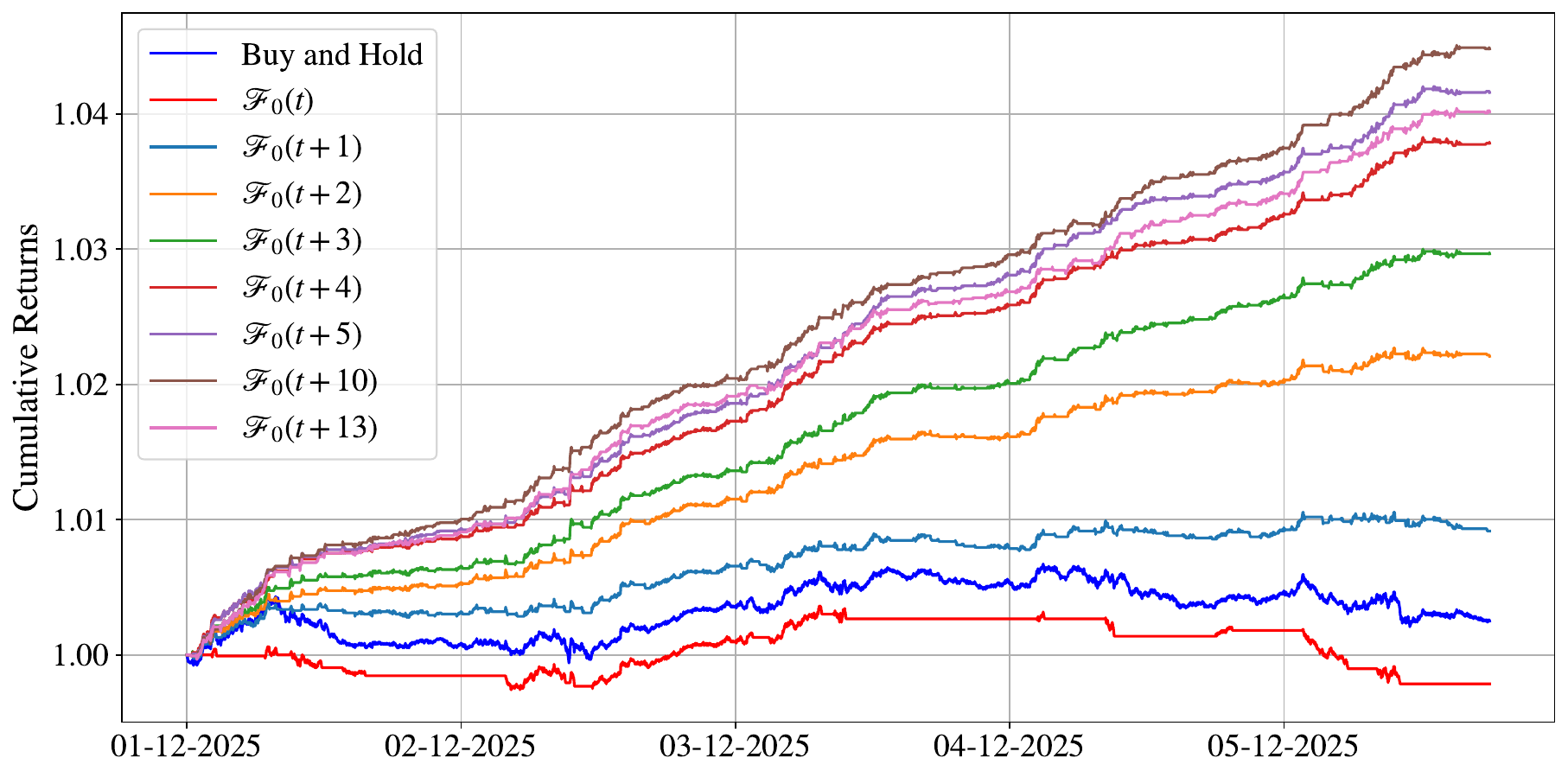}
\caption{Cumulative returns for the raw signal $\mathscr{F}_0(t)$ and its forward-shifted versions, illustrating that advancing the signal in time produces larger cumulative gains.}
\label{fig:target_pnl_example}
\end{figure*}

The raw composite signal $\mathscr{F}_0(t)$ is fully causal and constructed exclusively from information available up to time $t$.
While $\mathscr{F}_0(t)$ captures contemporaneous structure, it does not necessarily yield economically meaningful performance when mapped directly into positions over short horizons.
Figure~\ref{fig:target_pnl_example} illustrates this behavior in a representative interval, where $\mathscr{F}_0(t)$ underperforms compared to forward-shifted versions.

Motivated by this empirical observation, we introduce a causal transformation that injects forward-oriented structure while preserving online computability.
Specifically, we define
\begin{equation}
\mathscr{F}(t) = c_1(t)\,\mathscr{F}_0(t) + 2c_2(t)\,\widetilde{\partial_t \mathscr{F}_0}(t),
\end{equation}
where $\widetilde{\partial_t \mathscr{F}_0}(t)$ denotes a smoothed causal approximation
(moving average with span 4) of the first derivative $\mathscr{F}'_0(t)$.
In practice, the derivative is computed via a causal finite-difference estimate,
followed by a short-span moving-average smoothing (span equal to four observations),
which mitigates high-frequency oscillations while preserving local anticipatory structure. The numerical factor $2$ multiplying the derivative term is a fixed scaling constant,
introduced to balance the typical magnitudes of $\mathscr{F}_0(t)$ and
$\widetilde{\partial_t \mathscr{F}_0}(t)$ and kept constant throughout the analysis.

The inclusion of a derivative-based term is motivated by its ability to encode local forward tendency.
In smooth dynamical systems, the first derivative provides local information about the direction of evolution; for a sinusoid, the derivative corresponds to a phase-shifted signal.
While financial time series are neither deterministic nor periodic,
this analogy motivates the use of derivative information as a local
forward proxy under causality constraints, particularly in transition
regions where $\mathscr{F}_0(t)$ fluctuates around zero and directional
information becomes most relevant.

The mixing coefficients $c_1(t)$ and $c_2(t)$ are made state-dependent through $\mathscr{F}_0(t)$.
When $|\mathscr{F}_0(t)|$ is large, indicating a persistent regime, greater weight is assigned to $\mathscr{F}_0(t)$ itself.
When $\mathscr{F}_0(t)$ is near zero, derivative information becomes relatively more informative.
Accordingly,
\begin{equation}
c_1(t) = \tanh\!\left( \lvert \mathscr{F}_0(t) \rvert \right),
\qquad
c_2(t) = 1 - \tanh\!\left( \left\lvert \frac{\mathscr{F}_0(t)}{2} \right\rvert \right).
\end{equation}

Given the guiding observable $\mathscr{F}(t)$, the remaining step is to specify how it is translated into an actionable decision functional. We therefore proceed to define the decision rule and state update mechanism used in the empirical analysis.

\subsection{Decision Rule and State Update}
\label{subsec:trading_rule}

Given the guiding signal $\mathscr{F}(t)$, we specify a transparent state-based decision rule with a neutral zone controlled by a threshold $\theta > 0$.

Let $s_t = \mathscr{F}(t)$ denote the signal value at time $t$.
The state variable $p_t \in \{0,1\}$ indicates whether the system is in the ``active'' (invested/long) state or in the ``inactive'' (flat) state, respectively.
State updates follow a two-state hysteresis rule:
\begin{equation}
p_t =
\begin{cases}
1, & \text{if } p_{t-1}=0 \text{ and } s_t > \theta,\\
0, & \text{if } p_{t-1}=1 \text{ and } s_t < -\theta,\\
p_{t-1}, & \text{otherwise}.
\end{cases}
\label{eq:state_machine}
\end{equation}
This prevents rapid switching when the signal fluctuates around zero and ensures that state changes occur only when the signal crosses sufficiently strong positive or negative levels.

To preserve strict causality in the evaluation, the realized state applied at time $t$ is $p_{t-1}$ (one-step delay).
Denoting the simple return of the underlying asset by
\begin{equation}
r_t = \frac{P_t - P_{t-1}}{P_{t-1}},
\end{equation}
the realized response is computed as
\begin{equation}
R_t = p_{t-1}\, r_t.
\label{eq:strategy_return}
\end{equation}

Cumulative performance is reported via the compounded return series
\begin{equation}
V_t = V_0 \prod_{\tau=1}^{t} (1 + R_\tau),
\end{equation}
with $V_0$ normalized to one.
The buy-and-hold benchmark is computed analogously using $r_t$ in place of $R_t$.

Trading activity is quantified by counting state changes. Define
\begin{equation}
\Delta p_t = |p_t - p_{t-1}|,
\end{equation}
and total activity up to time $t$ by $\sum_{\tau=1}^t \Delta p_\tau$.

% =========================================================
\section{Experimental Results}
\label{sec:results_eurusdt}

This section reports empirical results for the EURUSDT exchange rate at the 1-minute frequency.
Importantly, all results in this section are computed without transaction costs.
This isolates the intrinsic economic relevance of the proposed causal observable from market frictions; the implications of even minimal fees are discussed in the concluding section.

Prior to computing any indicators, the raw price series is filtered to retain only active Forex trading hours (local time).
Saturdays are removed entirely, as well as Sundays before 18:00 and Fridays after 18:00.
This avoids distortions caused by weekend closures and illiquid periods and ensures that all indicators entering $\mathscr{F}_0(t)$ and $\mathscr{F}(t)$ are computed from contiguous market activity.

\begin{figure*}[!ht]
\centering
\includegraphics[width=0.8\textwidth]{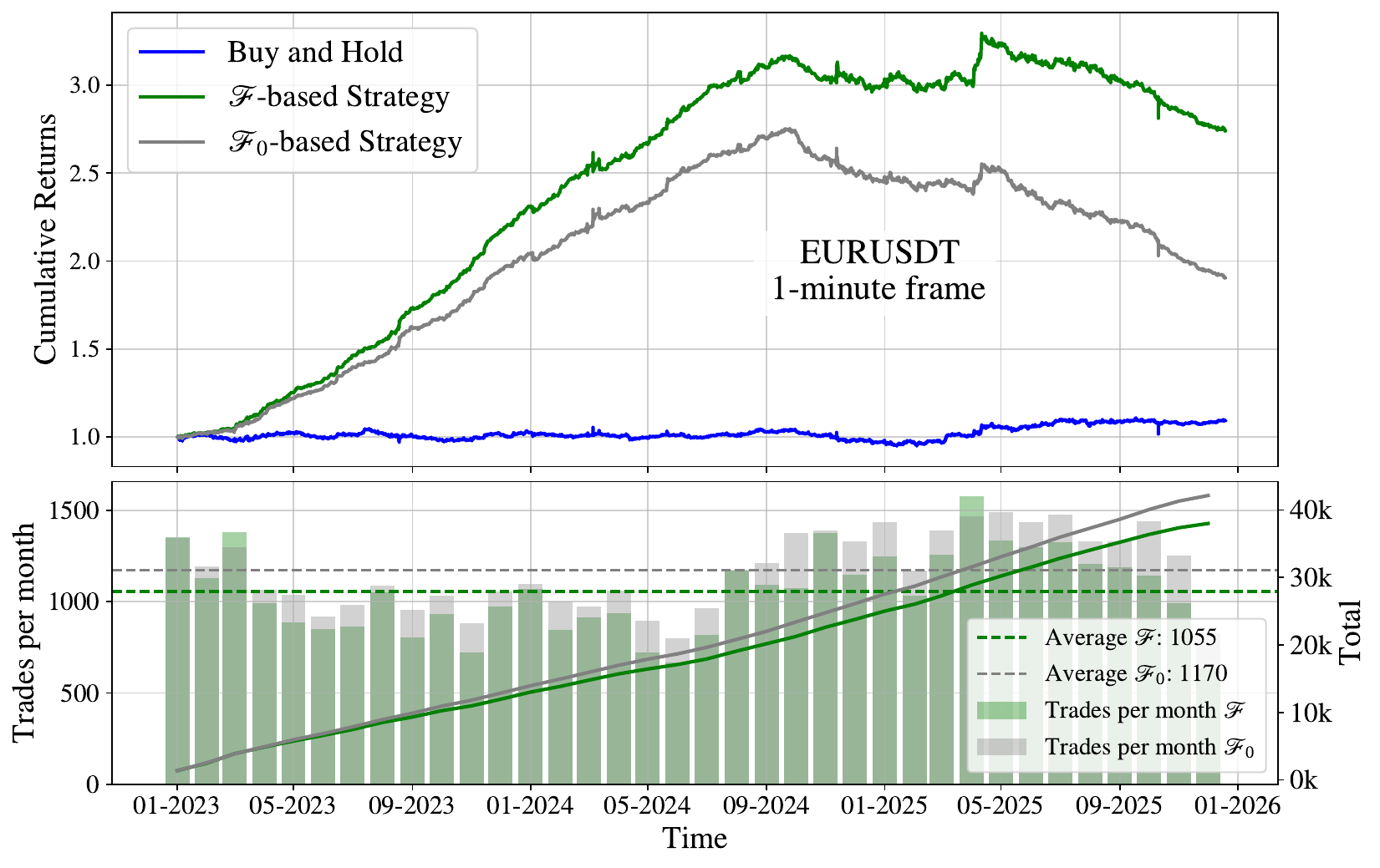}
\caption{Performance induced by applying the same threshold-based decision rule to the forward-oriented causal observable $\mathscr{F}(t)$ and to the raw composite signal $\mathscr{F}_0(t)$ for EURUSDT at the 1-minute frequency.
The upper panel shows cumulative returns relative to buy-and-hold, while the lower panel reports the cumulative number of state changes (trades).
The threshold is fixed at $\theta = 0.06$ for both signals.}

\label{fig:pnl_F_vs_bnh}
\end{figure*}

Figure~\ref{fig:pnl_F_vs_bnh} compares the cumulative returns induced by the proposed decision rule when applied to the forward-oriented causal observable $\mathscr{F}(t)$ and to the raw composite signal $\mathscr{F}_0(t)$, alongside a buy-and-hold benchmark.
The lower panel reports cumulative trading activity.
The threshold value $\theta = 0.06$ was chosen as a representative scale of the signal fluctuations, corresponding to a moderate percentile of the empirical distribution of $\mathscr{F}(t)$, and was kept fixed throughout the analysis; no parameter tuning or performance-based optimization was performed.

The figure reveals a pronounced regime dependence.
During the period from 2023 to approximately September~2024, the $\mathscr{F}$-based rule exhibits strong cumulative growth relative to buy-and-hold, whereas performance partially retraces and subsequently fluctuates around a near-plateau in the later period.
This behavior is quantified in Table~\ref{tab:regime_metrics}, which highlights the contrast between early growth and subsequent degradation, accompanied by sustained—and even increasing—trading activity in the second subperiod (around $10^3$ trades per month).
The persistence of high turnover despite diminishing gains indicates that the causal observable continues to trigger frequent state changes even as its incremental economic relevance deteriorates.

Importantly, Fig.~\ref{fig:pnl_F_vs_bnh} also shows that applying the same decision rule directly to the raw composite signal $\mathscr{F}_0(t)$ results in substantially weaker growth and earlier saturation.
While $\mathscr{F}_0(t)$ captures contemporaneous market structure, it lacks forward-oriented content.
This contrast demonstrates that the derivative-based component is essential: a purely smoothed or delayed version of $\mathscr{F}_0(t)$ does not introduce additional predictive structure, whereas the causal derivative contributes local directional information, particularly around regime transition regions.

Two empirical features are noteworthy. First, the induced dynamics exhibit substantial outperformance relative to buy-and-hold during extended intervals, most visibly throughout 2023 and up to approximately September~2024.
Second, from late 2024 onward, the cumulative performance partially retraces and then fluctuates around a near-plateau, while trading activity remains elevated. This combination of high turnover and limited incremental gains is consistent with the view that a fixed causal construction can become misaligned with the prevailing regime in a non-stationary environment.

\begin{table}[!t]
\centering
\caption{Incremental performance and activity metrics computed separately within each subperiod, with the cumulative return curve normalized to $V_0 = 1$ at the beginning of each regime. End equity $V$ denotes the final value.}
\label{tab:regime_metrics}
\resizebox{\columnwidth}{!}{%
\begin{tabular}{lrrrr}
\toprule
Period & End $V$ & Cum.\ ret.\ (\%) & MDD (\%) & Trades/mo \\
\midrule
2023--Sep 2024 & 3.14 & 214 & -4 & 959 \\
Post-Sep 2024  & 2.74 &  -13 & -17 & 1117 \\
\bottomrule
\end{tabular}}
\end{table}

% =========================================================
\section{Summary and Conclusion}
\label{sec:conclusion}

We proposed a transparent framework for constructing short-horizon predictive observables under strict causal constraints.
The methodology emphasizes signal engineering: (i) causal centering of heterogeneous indicator-level features, (ii) linear aggregation into a composite observable $\mathscr{F}_0(t)$, (iii) causal stabilization via a one-dimensional Kalman filter, and (iv) an adaptive forward-like operator that combines $\mathscr{F}_0(t)$ with a smoothed causal derivative term.
A key message is that economically relevant structure can be extracted without resorting to black-box predictors, provided causality is enforced and evaluation is disciplined.

In high-frequency EURUSDT, the induced dynamics show strong outperformance relative to buy-and-hold from 2023 through approximately September~2024, followed by a pronounced plateau thereafter.
This pattern is consistent with regime-dependent relevance in a non-stationary environment and motivates explicit mechanisms for regime adaptivity.
Importantly, results are intentionally reported without transaction costs to isolate intrinsic signal effects. Given the high turnover, even small frictions would materially reduce net performance, so the reported outcomes should be interpreted as evidence of predictive structure under idealized conditions rather than as an execution-ready trading strategy.

Future work naturally follows three directions: (i) friction-aware evaluation (spreads, slippage, latency) and turnover control; 
(ii) explicit regime adaptivity via change detection and time-varying thresholds or mixing functions $c_1(t),c_2(t)$; and 
(iii) generalizing the forward operator to a small causal basis of local features (multi-scale differences, short moving averages, or compact state-space augmentations), as well as a systematic investigation of the role of filtering choices and Kalman parameter sensitivity, while preserving interpretability and strict online computability.
These extensions keep the central philosophy intact: causal construction first, model complexity later.

% =========================================================
% =========================================================
 
% \bibliographystyle{elsarticle-harv}  
\bibliographystyle{elsarticle-num}
\bibliography{bib}

% =========================================================
% =========================================================

\appendix

\section{Technical Indicators}
\label{app:indicators}

This appendix documents the technical indicators employed in the construction.
All indicators rely exclusively on information available up to time~$t$, ensuring strict causality, n line with classical definitions of technical indicators in financial time series (see, e.g., \cite{murphy1999technical}).

\subsection{Relative Strength Index (RSI)}
The Relative Strength Index (RSI) is used as a bounded momentum measure and is defined \cite{wilder1978} as 
\begin{equation}
\text{RSI}_t = 100 \left( 1 - \frac{1}{1 + \frac{G_t}{L_t}} \right),
\end{equation}
where $G_t$ and $L_t$ denote average gains and losses, respectively.

\subsection{Money Flow Index (MFI)}
The Money Flow Index (MFI) is a volume-weighted momentum indicator defined as
\begin{equation}
\text{MFI}_t = 100 \left( 1 - \frac{1}{1 + \frac{\text{Positive Money Flow}}{\text{Negative Money Flow}}} \right).
\end{equation}

\subsection{Moving Average Convergence Divergence (MACD Difference)}
The MACD difference captures short-term trend acceleration \cite{appel1979}:
\begin{equation}
\text{MACD Diff}_t =
\bigl(\text{EMA}_{\text{fast}}(t) - \text{EMA}_{\text{slow}}(t)\bigr)
- \text{Signal}_t.
\end{equation}

\subsection{Bollinger Band Percent (BB\%)}
The Bollinger Band Percent (BB\%) provides a volatility-adjusted measure of price location:
\begin{equation}
\text{BB\%}_t =
\frac{P_t - (\mu_t - k\sigma_t)}{(\mu_t + k\sigma_t) - (\mu_t - k\sigma_t)}.
\end{equation}

\section{Kalman Filter}
\label{app:kalman}

To reduce high-frequency noise while preserving strict causality, the composite signal is smoothed using a one-dimensional Kalman filter \cite{kalman1960}.
The latent state follows a random-walk model,
\begin{equation}
x_t = x_{t-1} + w_t, \qquad w_t \sim \mathscr{N}(0,q),
\end{equation}
and observations are noisy measurements,
\begin{equation}
z_t = x_t + v_t, \qquad v_t \sim \mathscr{N}(0,r).
\end{equation}
The filter is implemented recursively using information available up to time~$t$ only, initialized with $x_0 = z_0$, and no backward smoothing is applied.

\end{document}